# VOLTAGE PERCOLATION THRESHOLDS EVIDENCED IN THE ELECTRICAL BEHAVIOUR OF DIFFERENT NANOSTRUCTURES

I. STAVARACHE[*],

*National Institute of Materials Physics, 105 bis Atomistilor Street, Magurele 077125, Romania*

Percolation phenomena are investigated and discussed in three kinds of nanostructures: first two are nanocrystalline silicon-based systems, Si nanodots embedded in amorphous $SiO_2$ matrix and porous silicon formed by an oxidized nanowire network, and the third consisting of a multi-walled carbon nanotube network embedded in amorphous SiN. The current-voltage characteristics measured on first two systems present voltage percolation thresholds with the same shape – a saturation plateau region of the current, followed by an abrupt increase. The current-voltage and conductance-voltage curves measured on multi-walled carbon nanotube network embedded in amorphous SiN present non-periodic and temperature independent oscillations. These oscillations are interpreted as voltage percolation thresholds.

*Keywords:* Percolation; Electrical transport; Nanocrystalline silicon; Carbon nanotubes.

## 1. Introduction

Percolation phenomena and their contribution to the physical properties of nanostructures have generated considerable research interest and intensive studies. Three forms of the basic percolation theory are known, namely site percolation theory, bond and continuum ones. The site percolation theory is commonly used in understanding of different characteristics, the bond percolation theory has applications in conduction processes in electronic devices and the continuum percolation theory is usually applied in porous media [1]. Transport properties are frequently described by means of the site and/or bond percolation theory. Let's imagine an infinite large box filled with metal (resistance $R$) and plastic balls (infinite resistance) of identical diameter. One finds the linear dimension ($\chi$) of the largest clusters (of interconnected metal balls) of the $\chi \propto \chi_0 |x - x_c|^{-\nu}$ form. $\chi_0$ represents a fundamental scale factor (the size of the ball), $x$ the fraction of metallic balls, $x_c$ the critical fraction of metallic balls and $\nu = 0.88$ (universal critical index of correlation radius [2]) in every three dimensional system [1]. If $x > x_c$, one can write the conductivity ($\sigma$) of an infinite system in three dimensions (3 D) in the $\sigma \propto (x - x_c)^2$ form. The numerical value of $x_c$ depends on the arrangement of the balls which are either random distributed or packed in some regular fashion, and it is different for site and bond percolation [1, 3].

The percolation phenomenon in disordered (semi)conducting systems can explain the carrier transport by choosing optimum conduction paths. If one considers a disordered system of nanocrystals uniformly dispersed in an amorphous matrix, a charge current can be measured if the concentration of nanocrystals is high enough to allow the carrier tunnelling. In other words, this system is a percolative one if the concentration is bigger than a critical value, $x_c$. At a percolation site the carrier chooses a road or another one, depending on the potential barrier met by it. The potential barrier can be tunnelled or hopped leading to a conduction process. In literature, the percolation phenomena in concentration as well as the concentration dependence of the conductance are frequently studied, in a wide range of materials.

In this paper, three types of nanostructures are discussed. One consists of Si dots embedded in amorphous $SiO_2$ (a-$SiO_2$) matrix (Si-$SiO_2$) and the second is formed by nanocrystalline porous silicon (nc-PS) with an oxidized nanowire network. MWCNT-SiN systems of multi-walled carbon nanotube (MWCNT) network embedded in amorphous SiN were also studied.

---

[*] Corresponding author: stavarache@infim.ro

Transport and optical properties related to structural ones of Si-$SiO_2$ systems were intensively investigated in the scientific literature [4, 5]. The microstructure of Si-$SiO_2$ sputtered films was described as being formed by Si nanocrystals (nc-Si), noncrystalline Si and $SiO_2$ islands [4−6]. Balberg *et al.* studied electrical transport and photoluminescence properties of films with nc-Si volume concentration ($x$) higher than percolation threshold concentration ($x_c$), equal or less than $x_c$ [5]. The dominant transport behaviour can be resumed briefly as migration bellow $x_c$, Coulomb blockade at $x_c$ and tunnelling above $x_c$. The curves of concentration dependence of the dark conductivity and photoconductivity were well fitted by power law $\sigma \propto (x - x_c)^t$, with $t \approx 2$, typical for percolation behaviour. Titova *et al* used time-resolved terahertz spectroscopy to investigate ultrafast carrier conduction in films of nc-Si embedded in an insulating $SiO_2$ matrix with various nc-Si content and sizes [7]. The extrapolated zero-frequency conductivity $\sigma_0$ was found to follow the $\sigma_0 \propto (x - x_c)^\gamma$ scaling law. They obtained $x_c \approx 38$ % and $\gamma \approx 1$.

The nc-PS layers are the most studied materials due to their structural, optical, dielectric and transport properties [8−10] and also due to their potential applications. The nc-PS film is often described as a three-component material, namely nanocrystalline Si that forms nanowire network, $SiO_x$ and pores (micro and/or nano) [2] or as a high porosity medium with a fractal geometry, a surrounding tissue and a distribution of nc-Si [10]. In the Ref. [2] the authors annalized analytically the dependence of PS conductivity on material porosity. They described the conductivity of PS films as the sum of two main electrical conduction mechanisms, i.e. crystalline and hopping processes. At low porosities the conductivity is mainly crystalline (on extended states). For porosities near percolation threshold PS is a percolation cluster, a mass fractal with fractal dimension of about 2.5. With the increasing of the porosity the 3D PS system is described as 1D system corresponding to silicon skeleton. Urbach *et al.* [10] proposed two main conduction mechanisms. The main conduction process in nonoxidized PS takes place via the disorder tissue surrounding the nc-Si, while in oxidized PS a hopping mechanism between nanocrystals is proposed. If the nc-Si skeleton is less oxidized so that the Si skeleton is not interrupted a conduction mechanism on extended states also takes place [8, 11, 12].

The electrical conductivity $\sigma$ of the CNT-based composites can be tailored directly by controlling the CNT content [13, 14], modulated by the CNT type (SWCNT, MWCNT), dimensions, purity. MWCNTs are reported in literature to be always electrically conductive even at low temperatures. Both SWCNTs and MWCNTs can have metallic or semiconductor character in electrical properties. The percolative behaviour of MWCNT-based systems was evidenced by the identification of a MWCNT critical concentration ($x_c$) or mass fraction ($f_c$) from the sharp increase of the conductivity when the MWCNT concentration (volume one - $x_{MWCNT}$ or mass fraction - $f_{MWCNT}$) becomes close to $x_c$ (or $f_c$) [14, 15]. When the MWCNT concentration reaches $x_c$ value a conductive path (network) is formed between the (measurement) electrodes. A power dependence of $\sigma$ on the MWCNT concentration, $\sigma \propto (x_{MWCNT} - x_c)^t$, where $t$ is a critical exponent, proved also the percolation phenomena and was evidenced in Ref. [13−15]. The values of $x_c$ ($f_c$) and $\sigma$ reported in literature present dispersion intervals depending on sizes or if CNTs are single, multi-walled or even bundles.

Ahmad *et al.* investigated the dc electrical conductivity and dielectric properties of the MWCNT-alumina composites at room temperature (RT) [14]. They showed that the electrical conductivity sharply increases with about eight orders of magnitude when MWCNTs concentration is close to percolation threshold of 0.79 vol %. For $x_{MWCNT} < x_c$ $\sigma = \sigma_i$ ($x_c$ − $x_{MWCNT})^{-s'}$, while for $x_{MWCNT} > x_c$ $\sigma = \sigma_c (x_{MWCNT} - x_c)^t$, where $\sigma_i$ and $\sigma_c$ are the dc conductivities of the insulating and conducting components, respectively. The $s'$ and $t$ exponents represent the critical exponents also corresponding to the insulating and conducting components. The exponent $t$ has a value between the theoretical ones corresponding to 2D (1.33) and 3D (1.94) [16].

MWCNTs in poly(*m*-phenylenevinylene-*co*-2,5-dioctyloxy-p-phenylenevinylene) and MWCNT-polyvinylalcohol composites were investigated as well [13]. The structures present an ohmic behaviour. From the dependence of $\sigma$ on the MWCNT mass fraction were found $f_c = 0.055$ wt % ($x_c \approx 0.029$ vol %) and $t \approx 1.36$. A fluctuation induced tunnelling mechanism was considered, taking into account that carriers tunnel through potential barriers of different heights due to local temperature fluctuations.

MWCNT-$Si_3N_4$ composites sintered by spark-plasma were investigated by González-Julián *et al* [15]. Dc and ac measurements and also conductive scanning force microscopy (C-SFM) investigations were performed. They evidenced bent and twisted MWCNTs at $Si_3N_4$ grain boundaries and also MWCNTs bundles. Therefore the electrical behaviour of structures is expected to be controlled by the nanotube network. Thus, a $\sigma$ increase of more than ten orders of magnitude was observed when the MWCNT concentration is increased from 0 to 0.9 vol %. The scaling law of $\sigma$ with $x_{MWCNT}$ was evidenced, with $t \approx 1.73$ at $x_c \approx 0.64$ vol %. The C-SFM investigations showed that the current flows only at the outermost layer of the nanotubes.

Gau *et al.* showed that the resistivity of hot-pressed MWCNT-polyamide-6 nanocomposites is not governed by the interconnected MWCNTs, but it is given by the polymer present between neighbouring MWCNTs [17]. The samples are composed of fibrils homogeneously distributed in the polymer. The authors proposed a conduction model in which electrons tunnel one by one through the polymer between the interconnected MWCNTs, forming a pathway. A decrease of resistivity with about two orders of magnitude is observed when $f_{MWCNT}$ varies between 0.75 and 4 wt %, and also when the temperature increases. The creation of an interconnecting conductive pathway was infirmed. The authors explained that the resistivity of the composite is dominated by the polymer, between the MWCNTs, one. The tunnelling current of the Wentzel-Kramers Brillouin form was considered.

In this paper we present results and discussions concerning the voltage percolation thresholds in different nanostructures. Three types of structures were investigated, i.e. nanocomposite films consisting of nanocrystalline Si dots embedded in amorphous $SiO_2$ matrix, nanocrystalline porous silicon, and multi-walled carbon nanotube networks embedded in amorphous SiN, respectively. All these structures have a percolative character. The voltage percolation thresholds were evidenced by the appearance of similar features at different voltages in the current-voltage ($I - V$) characteristics.

### 2. Description of nanostructures

The Si-$SiO_2$ nanocomposite films, formed of nanocrystalline Si dots embedded in an amorphous $SiO_2$ matrix, were obtained by co-sputtering, of Si and $SiO_2$ on quartz substrates [4, 6, 18], followed by a temperature annealing in $N_2$ at 1100 $^o$C. This way, samples with variable nc-Si volume concentration are obtained (from $x \approx 0$ % to $x \approx 100$ %), while the mean nanodot diameters vary slowly with $x$. After deposition of 50 parallel electrodes in a planar geometry, electrical measurements were performed between any two neighbour electrodes. These samples formed between neighbour electrodes have different nc-Si concentrations. High resolution transmission electron microscopy (HRTEM) images taken from the region with $x \approx 80$ % nc-Si reveal crystalline Si nanodots embedded in an amorphous $SiO_2$ matrix (see Fig. 1). These nanodots tend to form chains separated by a-$SiO_2$. The Si – $SiO_2$ structure is a percolative system [19], as it results from microstructure investigations and therefore the electrical transport governed by percolation processes is to be expected.

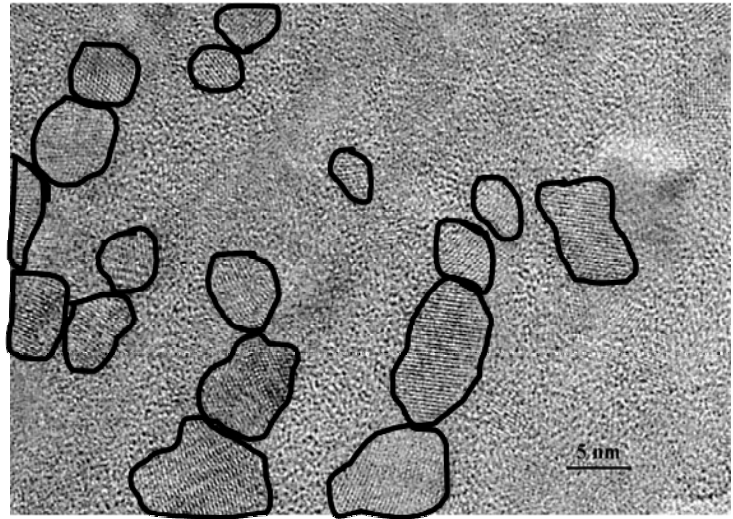

Fig. 1. *HRTEM image of a Si-SiO$_2$ sample with 80 vol % nc-Si.*

Nc-PS films were prepared by electrochemical etching of p-type Si wafers, (100)-oriented, with 5 – 15 Ω·cm resistivity, in HF (49%)-C$_2$H$_5$OH (1:1) electrolyte [8, 20]. Afterward, a photochemical etching was performed *in situ* by illumination with a Xenon lamp. A controlled oxidation also by anodization was carried out for the passivation of the internal surface. These films exhibit double scale porosity [8, 11]. The nc-PS layers present a macroporosity of 60 – 80 %. They are formed by a system of alveolar columnar macropores with a diameter of 1.5 – 3 μm, almost orthogonal on the PS film surface. These macropores cross the whole film thickness. The nanoporosity (≈ 50 %) is given by a nanowire network (1 – 5 nm diameter) that forms the alveolar walls (100 – 200 nm thickness). HRTEM image presented in Fig. 2, one observes the filament-like structure of the crystallized silicon (skeleton).

Other TEM images showed a clear separation between the crystallized skeleton and the amorphous SiO$_x$ one. The microstructure investigations show that nc-PS films represent a percolation network similar with that of Si-SiO$_2$ composites.

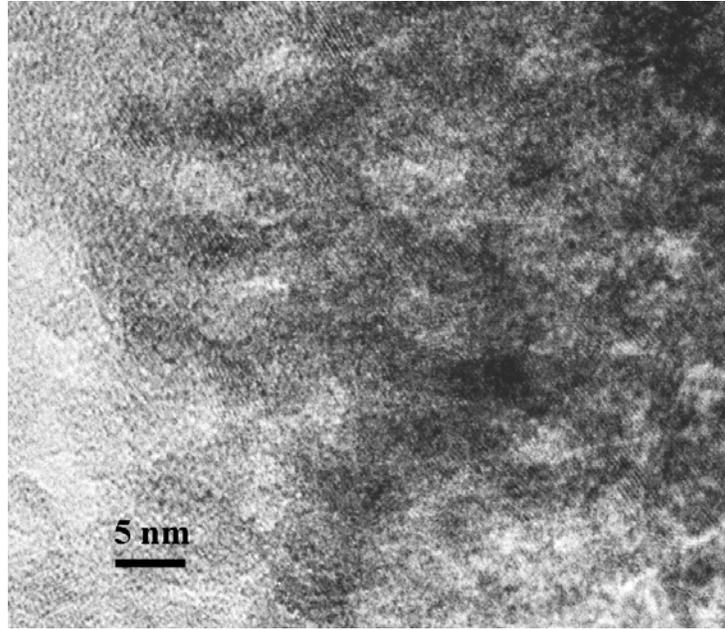

Fig. 2. *HRTEM image of an nc-PS fragment extracted from the film surface.*

The MWCNT-SiN structures were prepared in a sandwich geometry, i.e. quartz/Cr/Al/MWCNT-SiN/Cr/Al [21, 22], as one can see in the sketch illustrated in Fig. 3. The MWCNTs were embedded in SiN deposited by plasma enhanced chemical vapour deposition. Then, the SiN layer was etched in order to uncover the ends of MWCNTs. The cross-sectional transmission electron microscopy evidenced a fibrous morphology. It was demonstrated that MWCNT network is uniformly and homogeneously embedded in the amorphous SiN matrix [22].

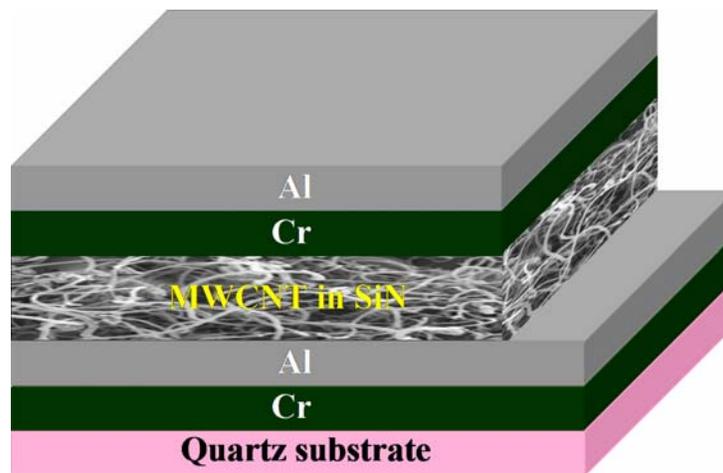

Fig. 3. *Quartz/Cr/Al/MWCNT-SiN/Cr/Al sandwich structure.*

### 3. Results and discussion

Figs. 4 and 5 illustrate current-voltage characteristics taken on the Si-SiO$_2$ films with two different nc-Si concentrations. If one analyses $I - V$ curve from Fig. 4 measured on a sample with 45.9 % nc-Si, several percolation thresholds that appear at 18.5, 21.5 and 23.5 V are evidenced. All these voltage thresholds have a similar shape, consisting of a saturation plateau region, followed by an abrupt current increase. This shape is independent on the nc-Si concentration as one can see also in the $I – V$ curve taken on a sample with $x ≈ 78.5$ % shown in Fig. 5. The voltage percolation thresholds, marked by arrows, are located at 2.7, 4.4, 13.4, 14.9, 19.5, 21.5, and 22.5 V.

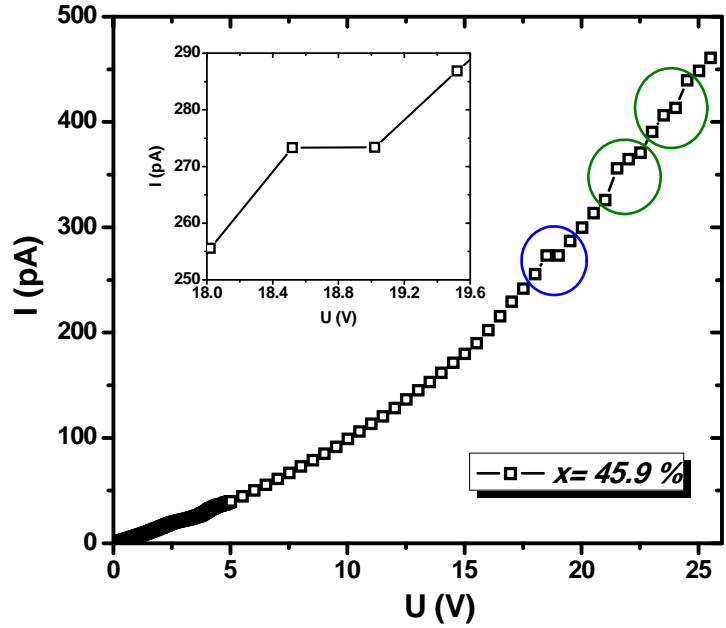

Fig. 4. *I – V characteristic for Si-SiO$_2$ sample with 45.9 % nc-Si. The percolation thresholds are marked. The blue one is shown in insert.*

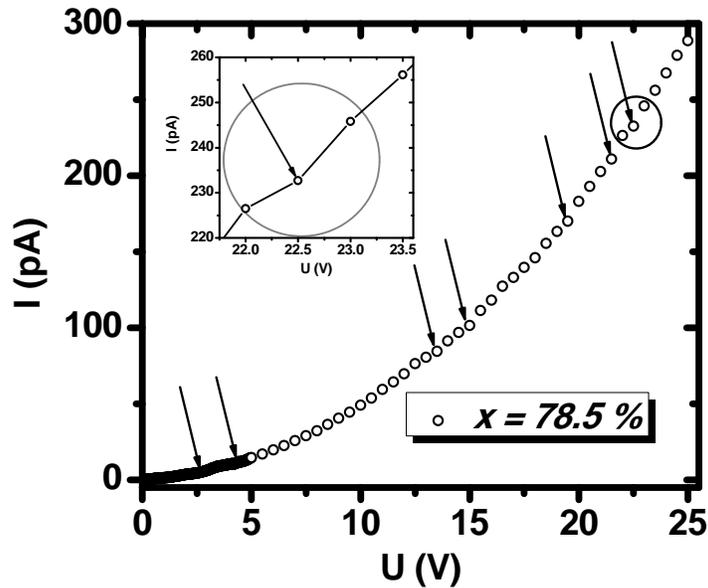

Fig. 5. *I – V curve taken on a Si-SiO$_2$ sample with $x ≈ 78.5$ % nc-Si. The voltage thresholds are marked with arrows. The encircled one is shown in insert.*

The percolative behaviour of Si-SiO$_2$ films as a function of $x$ was evidenced by the identification of the critical concentration $x_c$, i.e. concentration percolation threshold. Several measurements performed at low concentration allowed to identify the critical concentration corresponding to percolation threshold as $x_c \approx 34.7$ % [23]. Indeed, at lower concentration $x \approx$ 33.7 %, the current up to 2 V bias was less than or equal to the noise, at the same time as at higher $x \approx 35.7$ % the current was one order of magnitude greater [23]. The bias value of 2 V is a covering value, depending on the Coulomb blockade threshold. The estimated value of the Coulomb blockade threshold at RT is 1.4 V in the sample with $x \approx 34.7$ %. This value was determined by taking into account the value of 1.75 V calculated for films with $x \approx 66$ % [4]. The Coulomb blockade threshold is approximately proportional with the cubic root of the nc-Si concentration. Near $x_c$ the $I - V$ curves present a linear behaviour [23]. Similar values of the concentration percolation threshold in Si-SiO$_2$ systems were reported in literature. In the Ref. [5, 7] the reported value of the concentration percolation threshold is $x_c \approx 38$ %.

When increasing the nc-Si concentration, the current value increases too and the $I - V$ characteristics become superlinear. For example, the current measured on sample with $x \approx 43.9$ % is two orders of magnitude higher than that corresponding to $x \approx 35.7$ % [23]. The transport mechanism becomes the high field-assisted tunnelling, as results from the $I - V$ curves [4]. The carriers preferentially move along the nc-Si chains and tunnel the a-SiO$_2$ matrix, which separates the neighbouring nanocrystals. The current is well fitted by the equation corresponding to the high field-assisted tunnelling mechanism [4, 24], as follows:

$$I = I_0 \mathrm{sign}(V)\left[(1-|V|/V_0)\times\exp\left(-\alpha\sqrt{1-|V|/V_0}\right) - \exp(-\alpha)\right], \quad (1)$$

where parameters $I_0$, $V_0$ and $\alpha$ satisfy the equations: $I_0 = |a|\varphi$, $V_0 = N\varphi/e$, $\alpha = \delta\chi\varphi^{1/2}$ and $\chi$ is a constant equal to $(8m^*/\hbar^2)^{1/2}$. $a$ is a constant proportional to the number of equivalent paths, $\varphi$ is the mean height of the tunnelled barrier between nanocrystals, $N$ represents the mean number of barriers, $e$ is the elementary charge, $\delta$ is the mean width of the potential barrier between nanocrystals, and $m^*$ is the effective carrier mass inside the nanocrystal [4, 24]. The value of the potential barrier height, $\varphi = (2.2 \pm 0.2)$ eV, was experimentally determined from $I - V$ curves recorded on oxidized nc-PS [20]. Fig. 6 illustrates the $I - V$ characteristic taken on a sample with 78.5 % nc-Si, for both bias polarities, together with the theoretical fit given by Eq. (1). The fit parameters are $I_0 = 2.8731$ A, $V_0 = 140$ V and $\alpha = 25$, with a correlation coefficient of 0.9991. Thus, one obtains $N = 64$ tunnelled barriers with a mean width of 1.7 nm. The number of tunnelled barriers is small because with the increase of $x$, the number of equivalent paths with lower resistance sharply increases. In other words, one has to keep in mind that the estimated number of tunnelled barriers does not represent the total number of barriers existing in the sample between the electrodes. Some of barriers are short-circuited by the electrons that find less resistive paths.

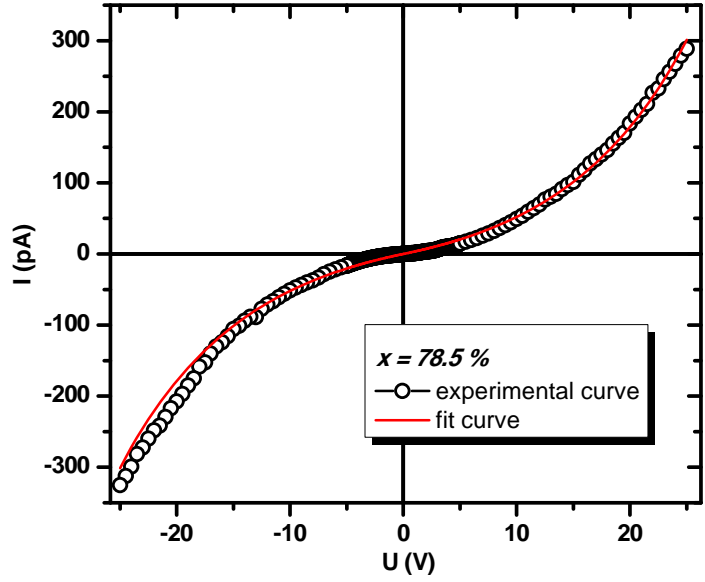

Fig. 6. *I – V curve for Si-SiO$_2$ sample with x ≈ 78.5 % nc-Si at both bias polarities. The red solid line represents the theoretical fit.*

The voltage percolation thresholds, evidenced in Figs. 4 and 5, are related to the random distribution of the nanocrystals, so that, at each voltage threshold, new (less resistive) paths are opened for the carriers. Thus, the increase of the applied voltage opens new paths by tunnelling larger barriers (e.g. if the barriers have a trapezoidal shape) and therefore the current abruptly increases. At high enough nc-Si concentration, all available paths are opened and Eq. (1) becomes rigorous.

Generally, in a percolative system the conductance depends on the concentration by a power law if the current-voltage behaviour is ohmic. In the case of Si-SiO$_2$ system, the $I - V$ curve is superlinear, so that the initial differential conductance ($G_0$), defined as $G_0 = (dI/dV)_{V=0}$, was found to follow a power scaling law, $G_0 \propto (x-x_c)^\gamma$. The critical exponent $\gamma$ has a value of 0.88 [23]. Titova *et al.* showed that the zero-frequency conductivity $\sigma_0$ also follows the $\sigma_0 \propto (x-x_c)^\gamma$ scaling law with $\gamma \approx 1$ [7]. In literature one finds $G \propto (x-x_c)^t$ with $t \geq 2 \neq \gamma$ [5]. The conductivity of an infinite system in 3D is $\sigma \propto (x-x_c)^2$ [1].

From Eq. (1) one obtains:

$$G_0 = \frac{I_0}{U_0}\left(\frac{\alpha}{2}-1\right)\exp(-\alpha). \qquad (2)$$

Exp($-\alpha$) is proportional with the tunnelling probability ($P_t$) that is proportional with the percolation probability ($P_p$), so that,

$$G_0 \propto P_t \propto P_p \propto (x-x_c)^\nu. \qquad (3)$$

One can see that the critical exponent $\gamma$ for the initial differential conductance is related to the tunnelling probability and is identified with the one for percolation probability ($\nu$). In the literature the exponent corresponding to a tunnelling process is reported to be $\nu \approx 0.88$, in good agreement with $\gamma \approx 0.88$. The value found for the critical exponent is close to the critical exponent previously found for nc-PS ($\gamma \approx 0.88$) [25].

Similar aspects were obtained for stabilized nc-PS films [26]. A typical $I - V$ characteristic measured at forward bias is displayed in Fig. 7. This $I - V$ curve presents two regions. One is exponential up to 2.2 V, while the second one is linear for the rest of the voltage interval. The value of 2.2 V represents the mean height of the potential barrier between the Si core (skeleton) of the nanowire and the $SiO_x$ shell [8]. Inspection of Fig. 7 leads also to the identification of several percolation thresholds in the quasi-linear region, located at 4.2, 7.2, 8.5, 9.3, 19.0, and 24.0 V. As one can observe, the voltage thresholds have the same shape as the ones evidenced in the $Si-SiO_2$ nanocomposites, namely a saturation plateau region of the current, followed by an abrupt increase.

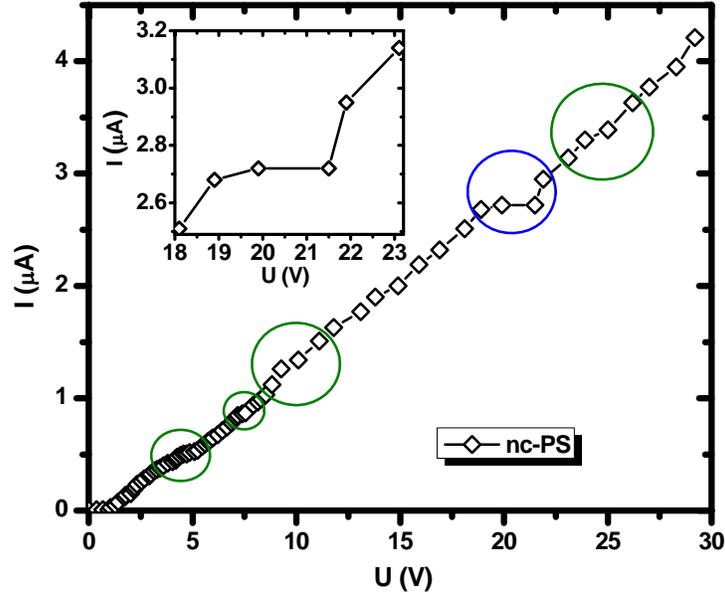

Fig. 7. *Forward I – V characteristic measured on nc-PS stabilized samples under forward bias. The percolation thresholds are marked. The blue one is shown in insert.*

In the nc-PS network, two types of conduction paths are present. The main conduction paths are along nanowires, but there are conduction paths through the oxide shells of adjoining ones. The current flows mainly along the nanowires and therefore it has an ohmic character. On the other hand, the wires form an intricate lump [8] and therefore percolation sites are expected to exist [27]. Therefore, when increasing the voltage, thin oxide barriers between neighbouring nanowires are tunnelled, and the current increases exactly in the same way as it happens in the case of $Si-SiO_2$ systems.

In the case of MWCNT-SiN nanostructures, the $I-V$ characteristics were measured at two different temperatures, low temperature (20 K) and RT (298 K), as one can see in Fig. 8. $I-V$ dependence is linear for both curves that shows a metallic-like behaviour of these structures. The linear character is also observed for resistance-temperature dependence [22]. González-Julián *et al.* investigated $MWCNT-Si_3N_4$ composites by measuring $I-V$ characteristics at different temperatures between 323 and 573 K [15]. They evidenced both semiconductor and metallic-like behaviours, dependent on the MWCNTs content and measurement temperature. In samples with low $x_{MWCNT}$, non-linear $I-V$ curves were obtained, $\sigma$ increasing with $T$, thus evidencing semiconductor behaviour. Hopping or tunnelling conduction across/through potential barriers at the internanotube contacts and across atomic defects within the nanotubes if MWCNTs are twisted and bended, were proposed. In samples with high $x_{MWCNT}$, the $I-V$ curves were found to be linear up to 423 K and non-linear at higher $T$. A metallic-like conduction associated to charge transport along the nanotube shells was evidenced up to 373 K.

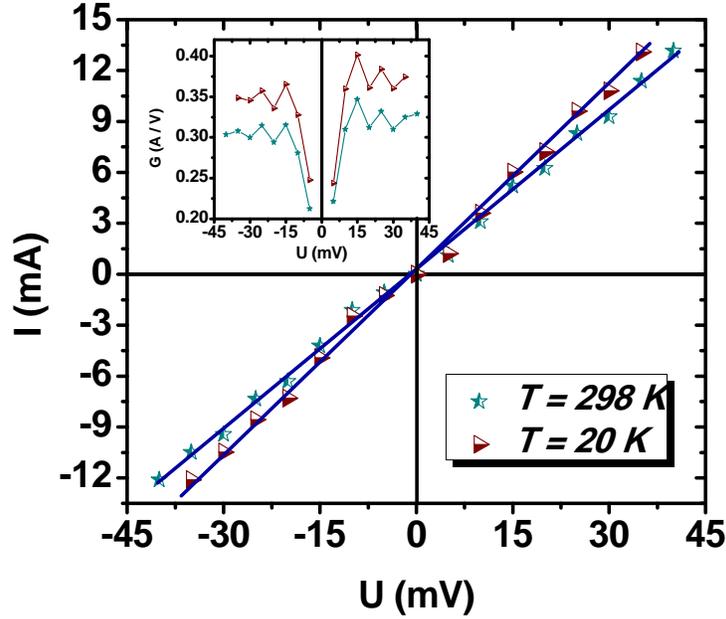

Fig. 8. *I – V curves taken on quartz/Cr/Al/MWCNT-SiN/Cr/Al samples at 20 and 298 K. The G – V curves are given in the insert.*

In our case, we have to remark small oscillations around the linear fit ($G \approx 0.31$ S for $T = 298$ K, $G \approx 0.36$ S for $T = 20$ K) on both $I – V$ characteristics. In order to evaluate these oscillations, the $G - V$ curves were represented (see the insert in Fig. 8). The conductance-voltage ($G - V$) dependences at both temperatures present maxima and minima symmetrical in bias polarization and located at the same voltages. Moreover, they are not periodic. These oscillations can not be attributed to the Coulomb blockade effect, since they are not temperature dependent and not periodic. We consider that they represent the signature of the percolation processes. The percolation in the disordered MWCNT network is due to the field-assisted tunnelling. The tunnelling takes place between adjacent MWCNTs embedded in the amorphous matrix, with a probability dependent on the relative orientation of the nanotubes and on the applied field. The minima and maxima sequence can be explained as follows. The current grows continuously with the voltage until it reaches a limit imposed by the quantification of conduction through the metallic nanotubes. This way, the $I – V$ curvature becomes sublinear and a conductance minimum is reached. Increasing the electric field up to a value representing the voltage percolation threshold, the tunnelling probability will increase, creating less resistive paths. The carriers will choose these paths, consequently the $I – V$ characteristic becomes superlinear and a maximum value of the conductance is reached. The voltage percolation thresholds of 20 and 30 mV on both bias polarities and both temperatures (20 and 298 K) correspond to the conductance minima. Gau *et al.* proposed in MWCNT-polymer composites a conduction model in which electrons tunnel one by one through the polymer between the interconnected MWCNTs, forming a pathway [17].

### 4. Conclusions

The $I – V$ characteristics measured on Si-SiO$_2$ nanocomposite films and nc-PS present voltage percolation thresholds with the same shape, namely a saturation plateau region of the current, followed by an abrupt increase. These thresholds appear in the percolative systems and they are due to tunnelling through potential barriers between neighbouring Si dots/Si wires. These potential barriers are given by the silicon oxides. In spite of the fact that the dominant transport mechanisms are different in both nanostructures, the percolation is always due to the tunnelling process. In conclusion, the macroscopic transport in silicon-based nanocrystalline systems with random space distribution of the nanocrystallites is percolative, whatever the microscopic transport mechanism may be.

In the Si-SiO$_2$ films, the critical exponent of 0.88 for initial differential conductance is identified with the one for tunnelling probability and also with the critical exponent for the percolation probability.

In the case of MWCNT network embedded in SiN, similarly with the silicon-based nanostructures, the oscillations displayed in the $I - V$ and $G - V$ curves are due to percolation processes too. These oscillations are symmetrical in bias polarization and are not periodic and are temperature independent. The voltage percolation thresholds of 20 and 30 mV on both bias polarities and both temperatures (20 and 298 K) correspond to the conductance minima.

The analysis of the voltage percolation thresholds in the $I - V$ and $G - V$ curves represents a novel and original approach to the study of the percolation phenomena in Si-SiO$_2$, nc-PS and MWCNT-SiN systems. It can be extended to any nanostructure containing random distribution of nanodots, nanowires or nanotubes embedded in an insulating and/or amorphous matrix.

**Acknowledgements**

This work was supported by the Romanian National Authority for Scientific Research through the CNCSIS –UEFISCDI Contract No. 471/2009 (ID 918/2008).**References**

[1] A. Hunt, Complexity **15**, 13 (2009).
[2] V. M. Aroutiounian, M. Zh. Ghulinyan, Phys. Stat. Sol. (a) **197**, 462 (2003).
[3] D. Stauffer, Classical Percolation, Lect. Notes. Phys. **762**, 1 (2009).
[4] M. L. Ciurea, V. S. Teodorescu, V. Iancu, I. Balberg, Chem. Phys. Lett. **423**, 225 (2006).
[5] I. Balberg, E. Savir, J. Jedrzejewski, A. G. Nassiopoulou, S. Gardelis, Phys. Rev. B **75**, 235329 (2007).
[6] V. S. Teodorescu, M. L. Ciurea, V. Iancu, M.-G. Blanchin, J. Mater. Res. **23**, 2990 (2008).
[7] L. V. Titova, T. L. Cocker, D. G. Cooke, X. Wang, Al Meldrum, F. A. Hegmann, Phys. Rev. B **83**, 085403 (2011).
[8] M. L. Ciurea, V. Iancu, V. S. Teodorescu, L. C. Nistor, M. G. Blanchin, J. Electrochem. Soc. **146,** 3516 (1999).
[9] M. L. Ciurea, M. Draghici, V. Iancu, M. Reshotko, I. Balberg, J. Luminesc. **102-103**, 492 (2003).
[10] B. Urbach, E. Axelrod, A. Sa'ar, Phys. Rev. **B** 75, 205330 (2007).
[11] M. L. Ciurea, V. Iancu, Proc. IEEE CN 00TH8486, Int. Semicon. Conf. CAS 2000, Vol. **1**, 55 (2000).
[12] R. Jemai, A. Alaya, O. Meskini, M. Nouiri, R. Mghaieth, K. Khirouni, S. Alaya, Mater. Sci. Eng. B **137**, 263 (2007).
[13] B. E. Kilbride, J. N. Coleman, J. Fraysse, P. Fournet, M. Cadek, A. Drury, S. Hutzler, S. Roth, W. J. Blau, J. Appl. Phys. **92**, 4024 (2002).
[14] K. Ahmad, W. Pan, S.-L. Shi, Appl. Phys. Lett. 89, 133122 (2006).
[15] J. González-Julián, Y. Iglesias, A. C. Caballero, M. Belmonte, L. Garzón, C. Ocal, P. Miranzo, M. I. Osendi, J. Compos. Sci. Technol. 71, 60 (2011).
[16] D. Stauffer, Introduction to the percolation theory. London and Philadelphia: Taylor & Francis (1985).
[17] C. Gau, C.-Y. Kuo, H. S. Ko, Nanotechnol. **20**, 395705 (2009).
[18] V. S. Teodorescu, M. L. Ciurea, V. Iancu, M. G. Blanchin, Proc. IEEE CN 04TH8748, Int. Semicond. Conf. CAS 2004, Vol. **1**, 59 (2004).- 11 -